# AI-based virtual patient interactions improve professional language proficiency in international nurses

## Abstract

**Objectives:** International healthcare personnel face significant challenges in language acquisition. We investigated whether virtual patient interaction (VI) training can support professional integration and clinical language proficiency.

**Methods:** In an explorative controlled study, 25 international nursing staff undergoing professional qualification measures in Germany were included. The intervention group (n=13) received access to casuu, a digital simulation tool providing interactive communication scenarios with virtual patients and colleagues. The control group (n=12) followed standard language training without VI. Both groups received 600 hours of training for the B2 for Nursing exam, as required for professional licensure in several German states, at the same language school using the same textbook. The primary outcome was the fail rate at the first attempt, with effects on different skill dimensions assessed in addition. The clinical supervisors were interviewed 12-14 months after the arrival.

**Results:** The fail rate in the B2 for Nursing exam was reduced from 66.7% in the control group to 25.0% in the intervention group (relative reduction 62.5%). Competence increased primarily in the skill dimension “active speaking”. Attrition rates did not increase (1/13 vs. 2/12). Clinical superiors reported increased confidence and competence in team and patient interactions in the intervention group.

**Conclusion:** Virtual interaction-based training improves the acquisition of professional language and clinical communication skills of international nurses. It integrates with conventional teaching and does not increase dropout rates.

**Keywords:** international healthcare personnel, professional language training, nursing education, virtual interaction training, virtual patients

**Sebastian Ullrich**[1]
**Jan-Marcus Haus**[2]
**Sandra Lamb**[3]
**Nicole Hermandung**[4]
**Tom Friemel**[5]
**Olav Stolze**[3]
**Sabine Schmidt**[3]
**Olha Demianiuk**[6]
**René Werner**[7,8,9]
**Parisa Moll-Khosrawi**[2]
**Rüdiger Schmitz**[6,7,8]

1 Städtisches Krankenhaus Kiel, Department of Internal Medicine III, Kiel, Germany

2 University Medical Center Hamburg-Eppendorf, Department of Anaesthesia, Hamburg, Germany

3 Städtisches Krankenhaus Kiel, Directorate of Nursing, Kiel, Germany

4 DRK Akademie SH, Kiel, Germany

5 Städtisches Krankenhaus Kiel, Department of Internal Medicine I, Kiel, Germany

6 casuu GmbH, Hamburg, Germany

7 University Medical Center Hamburg-Eppendorf, Institute for Applied Medical Informatics, Hamburg, Germany

8 University Medical Center Hamburg-Eppendorf, Institute for Computational Neuroscience, Hamburg, Germany

9 University Medical Center Hamburg-Eppendorf, Center for Biomedical Artificial Intelligence (bAIome), Hamburg, Germany

## Introduction

The shortage of skilled nursing personnel in Germany and many countries is acute and persistent. A central countermeasure is the integration of international nursing staff, with a recent net immigration of 36,000 persons in Germany (+36% to previous year [1]). Since 2021, the increase in employment within nursing professions has been entirely driven by immigration, while the number of nationals in these professions is actually declining [2].

Language barriers and lack of specific language training are repeatedly reported as the most important barriers in this process [3], [4], [5]. Limited language proficiency correlates with work-life balance problems [6] and stress [7] as early warning signals for job discontinuation [8]. As a result, 32.1% of employment contracts with internationally recruited nurses end within the first 12 months [9], implying limited language proficiency as a potentially relevant factor in job discontinuation. Among those who continue, high failure rates in Kenntnisprüfung (KP) and/or professional language examinations can delay the process by months until full professional recognition and clinical readiness are reached. The authors are aware of anecdotal evidence of first attempt failure rates of up to 90% for certain examinations and cohorts.

There is reason to assume that language barriers may adversely affect the quality of medical care – an effect that is, at minimum, well-documented from the patient's perspective [10]. The available data for the effect of language barriers on the side of the healthcare provider is limited. However, there is plenty of data that, reversing the situation and considering patients who are native in a foreign language, suggests that the effect of limited language proficiency on patient safety is substantial [11], [12], [13].

Efforts to facilitate a qualitative, quick and reliable (professional) language acquisition are thus of the utmost importance in times of labour shortage and international migration. This study evaluates the effect of a novel, AI-based virtual training scheme on the professional language proficiency of international nurses.

## Material and methods

A total of 25 internationally trained nursing staff undergoing professional qualification measures were enrolled (intervention group, n=13; control group, n=12). Both groups attended parallel 6 month, 600-hour qualification courses at the same language school (DRK Akademie Schleswig-Holstein, Kiel) and using the same textbook, ensuring consistency in teaching quality and curriculum. 92% of the participants (23/25) originated from the same home country, Tunisia, the others from Nepal (n=1) and Ukraine (n=1). All entered training with comparable initial German language proficiency (CEFR level B1) and comparable demographics (cf. table 1).

The intervention group received access to casuu (casuu GmbH, Hamburg, Germany, [https://casuu.com]), a digital simulation platform that provides interactive, AI-supported communication scenarios simulating real-life clinical situations with virtual patients and colleagues [14]. Usage was mostly self-directed but encouraged, with the exception of three approximately two-week phases of structured integration into the classroom-setting including assigned homework and integrated debriefing sessions. The 15-20 minute debriefing sessions involved individual students presenting the transcripts of their VIs to the class via a projector. Guided by the teacher, the class then discussed newly emerging lexical fields, vocabulary gaps, and whether the expressions used had already been covered in class. The control group attended standard language training that did not include VI training, but otherwise did not differ in terms of curriculum or contents.

In parallel to the language courses, participants in both groups underwent clinical onboarding on the wards of the Städtisches Krankenhaus Kiel (SKK). Over the course of the program, the ratio of clinical work to language training gradually shifted from approximately 1:4 days per week at the beginning to 4:1 toward the end. The structure and intensity of language teaching and practical clinical work were identical for both groups.

The primary outcome was the first-attempt failure rate at the B2 for Nursing professional language examination (telc® Deutsch B2 Pflege) as required for professional licensure in several German states. Both groups conducted this exam at the same, external, institution (Inlingua Sprachcenter Kiel, Kiel) approximately 8 months after arrival. The overall score at the exam was recorded as a secondary outcome. Furthermore, training effects across different language skill domains – receptive skills (reading and listening), writing and speaking – were analyzed using the anonymized section-level results from the same examination. Attrition rates from the qualification program were also recorded. The primary outcome was then analyzed both for all enrolled participants (intention-to-treat, ITT) and for those who continued until the language examination (per-protocol, PP).

To explore clinical instructors' perspective on the onboarding and integration of the nursing staff, semi-structured interviews were conducted with four clinical supervisors (Praxisanleiter*innen) at SKK. Each interview lasted between 30 and 45 minutes and was conducted by a single interviewer, who documented the interviews in structured summary notes. Prior to the interview, participants provided informed consent to anonymous processing of their responses. The summary of the interviews was then reviewed and verified by three independent readers to ensure accuracy and completeness. The thematic structure of the interview guide provided a deductive framework, within which the participants' responses were condensed into categories. The interview guide is available in the attachment 1.

**Table 1: Demographics, qualification level and professional experience**

Note that for the professional experience, years in nursing in a hospital setting were considered, and data rounded to full years.

| | Control | Intervention | p |
|---|---|---|---|
| Age [yrs]: mean (std. dev) | 30.2 (8.3) | 30.1 (5.1) | 0.97 |
| Gender [% male] | 23.1 | 33.3 | 0.57 |
| BA Nursing or equiv. [%] | 100 | 100 | |
| Experience [years*] | 1.23 (0.83) | 1.00 (0.00) | 0.35 |

## Results

Of those who attended the B2 for Nursing exam, 25.0% (3 out of 12) failed in the intervention group as compared to 66.7% (6 out of 9) in the control group, corresponding to a relative reduction of 62.5% (p=0.028) per protocol.

During the course of the training, 1 participant from the intervention group terminated their contract with SKK and dropped out of the training. In the control group, this was the case for two nurses. In addition, in the control group, 1 person took their B2 exam while on vacation at home, effectively violating the study protocol. To avoid overestimating the effect of the intervention, this case was conservatively counted as a pass. Therefore, in the ITT analysis, only 30.8% (4 out of 13) in the intervention group were counted as “failed” (failed at the exam or terminated early), as compared to 66.7% (8 out of 12) of those who received the traditional training alone. This corresponds to a relative reduction in the fail rate of 53.8% (p=0.036).

Overall exam scores were 140.25±17.15 (125.50±14.89) in the intervention (control) group (p=0.030). With only minor and insignificant differences in “receptive” and “writing” skill dimensions, the difference was mainly attributed to “active speaking” (arithmetic mean 88.58 ± standard deviation 8.83 versus 77.39±9.84, p=0.009).

The attrition rate within the qualification programs was 1/13 in the VI group vs. 2/12 in the control group, indicating that the additional training did not increase dropout risks. The intensity of the VI training was higher in phases when it was systematically integrated into the classroom-based course by the structured debriefings (62-100% weekly active users) as compared to other weeks (0-62%).

Qualitative feedback from individual supervisors suggested that VI-trained participants exhibited greater confidence in clinical communication and interpersonal interactions. In a round of semi-structured interviews of clinical supervisors, all four interviewees noted a marked reduction in language anxiety and an increase in communicative confidence. Nurses were consistently described as speaking more fluently, engaging more actively in team interactions, and performing more capably in patient-facing situations. All respondents drew a direct link between language competence and patient safety. Higher language proficiency in the intervention group, in the view of the clinical supervisors, also helped in facilitating a faster onboarding process as witnessed by the clinical supervisors.

## Conclusion and outlook

The results show that virtual interaction (VI) training can substantially improve professional language acquisition in international healthcare personnel. Integration of VI training did not increase dropout rates and was particularly effective on active speaking skills.

This study has a number of limitations: First, the overall sample size is limited. Second, a larger effect may be hypothesized if starting earlier and in the home country. Both will be the focus of a larger, government-funded study (V-PREP-CARE). Third, participants were not restricted from completing the exercises collaboratively, assumingly blurring data on user activity in phases without classroom-integrated usage. Fourth, although not actively informed, it can be assumed that the clinical supervisors (Praxisanleiter) were aware of which participants underwent additional VI training. The results from the semi-structured interviews are therefore potentially subject to observer bias. In contrast, the examiners of the B2 for Nursing examination were blinded to participants’ group allocation. Fifth, given the summarized character of the qualitative data, findings of the interviews reflect interpreted content rather than literal transcribed accounts of the interviewees. They should therefore be understood more as exploratory rather than exhaustive. Sixth, anonymized test results reported by the examination body distinguished only the skill dimensions “receptive”, “writing”, and “active speaking”, effectively merging listening and reading into a single category. Nevertheless, the data suggest that listening and reading comprehension may be particularly relevant targets for future refinements of the training program.

Virtual interaction training holds potential for reducing systemic costs and time until clinical readiness, enhancing the efficiency of workforce integration, and improving healthcare delivery in the context of labor shortage and workforce migration. It should be considered as a potentially important element of sustainable and future-ready strategies for addressing the nursing workforce shortage in Germany and beyond.

# Notes

## Author contributions

US and JMH share first authorship.
PMK and RS share last authorship.

## Authors' ORCIDs

- Jan-Marcus Haus: [0000-0003-0012-4137]
- Tom Friemel: [0009-0001-1250-9256]
- Olha Demianiuk: [0009-0000-6784-5305]
- René Werner: [0000-0002-2296-3305]
- Parisa Moll-Khosrawi: [0000-0003-2024-0020]
- Rüdiger Schmitz: [0000-0002-2414-9817]

# Acknowledgements

We are grateful towards the nurses who took part in this study, their teachers and staff at DRK Akademie SH, as well as their supervisors and staff at Städtisches Krankenhaus Kiel. We acknowledge support by Inlingua Kiel and telc® gGmbH. We thank Sebastian Zimmermann of telc gGmbH for the compilation and anonymization of the exam results.

# Competing interests

US, OS and SL have roles in the follow-up study V-PREP-CARE. RW is an advisor to casuu GmbH [https://casuu.com/] on a pro bono basis and without financial interests. RS is a co-founder and shareholder of casuu. OD is a student assistant to the company.
The other authors declare that they have no competing interests.

# Attachments

Available from https://doi.org/10.3205/zma001868
1. Attachment_1.pdf (216 KB)
   Interview guide: Clinical instructors and international nurses

**Corresponding author:**
Rüdiger Schmitz
University Medical Center Hamburg-Eppendorf, Institute for Applied Medical Informatics, Quartier am Zeughaus, Christoph-Probst-Weg 1, D-20251 Hamburg, Germany
r.schmitz@uke.de

# KI-basierte virtuelle Patient*inneninteraktionen verbessern die berufsbezogene Sprachkompetenz internationaler Pflegefachpersonen

## Zusammenfassung

**Ziele:** Internationales Gesundheitspersonal steht beim Spracherwerb vor erheblichen Herausforderungen. Wir untersuchten, ob ein Training mit virtuellen Patient:inneninteraktionen (VI) die berufliche Integration und den Erwerb der klinischen Sprachkompetenz unterstützen kann.
**Methoden:** Kontrollierte Studie mit 25 international ausgebildeten Pflegefachpersonen im Anerkennungsverfahren. Beide Gruppen absolvierten einen 600-stündigen Vorbereitungskurs für die Prüfung B2 Pflege an derselben Sprachschule und mit demselben Lehrbuch. Die Interventionsgruppe (n=13) erhielt zusätzlich Zugang zu casuu, einem digitalen Simulationstool, das interaktive Kommunikationsszenarien mit virtuellen Patient:innen und Kolleg:innen bereitstellt. Die Kontrollgruppe (n=12) absolvierte den Sprachkurs ohne ergänzendes VI. Primärer Endpunkt war die Durchfallquote beim ersten Prüfungsversuch; zusätzlich wurde die Wirkung des Trainings auf verschiedene Kompetenzdimensionen untersucht. Außerdem wurden die Praxisanleiter:innen der Teilnehmer:innen 12-14 Monate nach der Ankunft interviewt.
**Ergebnisse:** Die Durchfallquote in der Prüfung B2 Pflege sank von 66,7% in der Kontrollgruppe auf 25,0% in der Interventionsgruppe, was einer relativen Reduktion um 62,5% entspricht. Die Kompetenz nahm vor allem in der Dimension „aktives Sprechen“ zu. Die Abbruchquoten stiegen nicht an (1/13 vs. 2/12). Klinische Vorgesetzte berichteten von erhöhter Sicherheit und Kompetenz in Team- und Patient:inneninteraktionen in der Interventionsgruppe.
**Schlussfolgerung:** Ein auf virtuellen Interaktionen basierendes Training verbessert den Erwerb berufsbezogener Sprachkompetenz und klinischer Kommunikationsfähigkeiten bei internationalen Pflegefachpersonen. Es lässt sich in den konventionellen Unterricht im Rahmen des Anerkennungsverfahrens integrieren und erhöht die Abbruchquoten nicht.

**Schlüsselwörter:** Fachkräftemangel, Pflege, internationale Pflegekräfte, Zuwanderung, berufsbezogenes Sprachtraining, virtuelle Interaktionen, virtuelle Patient*innen

**Sebastian Ullrich**[1]
**Jan-Marcus Haus**[2]
**Sandra Lamb**[3]
**Nicole Hermandung**[4]
**Tom Friemel**[5]
**Olav Stolze**[3]
**Sabine Schmidt**[3]
**Olha Demianiuk**[6]
**René Werner**[7,8,9]
**Parisa Moll-Khosrawi**[2]
**Rüdiger Schmitz**[6,7,8]

1 Städtisches Krankenhaus Kiel, Klinik für Innere Medizin III, Kiel, Deutschland

2 Universitätsklinikum Hamburg-Eppendorf, Klinik und Poliklinik für Anästhesiologie, Hamburg, Deutschland

3 Städtisches Krankenhaus Kiel, Pflegedirektion, Kiel, Deutschland

4 DRK Akademie SH, Kiel, Deutschland

5 Städtisches Krankenhaus Kiel, Klinik für Innere Medizin I, Kiel, Deutschland

6 casuu GmbH, Hamburg, Deutschland

7 Universitätsklinikum Hamburg-Eppendorf, Institut für Angewandte Medizininformatik, Hamburg, Deutschland

8 Universitätsklinikum Hamburg-Eppendorf, Institut für Computational Neuroscience, Hamburg, Deutschland

9 Universitätsklinikum Hamburg-Eppendorf, Center for Biomedical Artificial Intelligence (bAIome), Hamburg, Deutschland

## Einleitung

Der Mangel an qualifiziertem Pflegepersonal ist in Deutschland und in vielen anderen Ländern akut und anhaltend. Eine zentrale Maßnahme zur Kompensation ist die Integration internationaler Pflegefachpersonen. In Deutschland lag die Nettozuwanderung in diesem Bereich jüngst bei etwa 36.000 Personen (+36% im Vergleich zum Vorjahr [1]). Der Beschäftigungszuwachs in Pflegeberufen insgesamt ist seit 2021 vollständig durch Zuwanderung getragen, während die Zahl der deutschen Staatsangehörigen in diesen Berufen tatsächlich rückläufig ist [2].
Berufsbezogene Sprachbarrieren und fehlende spezifische Sprachförderung werden in diesem Prozess wiederholt als die wichtigsten Hindernisse beschrieben [3], [4], [5]. Eine eingeschränkte Sprachkompetenz korreliert mit Problemen der Work-Life-Balance [6] und Stress [7] als frühen Warnsignalen für einen Abbruch der Berufstätigkeit [8]. 32,1% der Arbeitsverträge mit international angeworbenen Pflegefachpersonen enden innerhalb der ersten 12 Monate [9], was auf eine eingeschränkte Sprachkompetenz als potenziell relevanten Faktor für den Abbruch der Berufstätigkeit hinweist. Bei denjenigen, die im Beruf verbleiben, können hohe Durchfallquoten in der Kenntnisprüfung (KP) und/oder in berufsbezogenen Sprachprüfungen den Prozess bis zur vollständigen beruflichen Anerkennung und klinischen Einsatzfähigkeit um Monate verzögern. Den Autor*innen sind Durchfallquoten von bis zu 90% im Erstversuch bei einzelnen Kohorten und Prüfungen anekdotisch bekannt.
Es ist naheliegend, dass Sprachbarrieren auch die Qualität der medizinischen Versorgung beeinträchtigen können – eine Annahme, die zumindest aus der Patient*innenperspektive gut dokumentiert ist [10]. Die verfügbaren Daten zum Effekt von Sprachbarrieren aufseiten der Leistungserbringenden im Gesundheitswesen sind begrenzt. Es existieren jedoch zahlreiche Daten, die bei umgekehrter Betrachtung – also bei Patient*innen, deren Muttersprache eine andere Sprache ist – darauf hindeuten, dass die Auswirkungen eingeschränkter Sprachkompetenz auf die Patient*innensicherheit erheblich sind [11], [12], [13].
Bemühungen um und neue Ansätze für einen qualitativ hochwertigen, schnellen und verlässlichen Erwerb von (berufsbezogener) Sprachkompetenz sind daher in Zeiten von Arbeitskräftemangel und internationaler Migration von größter Bedeutung. Diese Studie untersucht daher den Effekt eines neuartigen, KI-basierten, virtuellen Trainingskonzepts auf die berufsbezogene Sprachkompetenz internationaler Pflegefachpersonen.

## Material und Methoden

25 international ausgebildete, examinierte Pflegefachpersonen im Anerkennungsverfahren wurden in die Studie eingeschlossen (Interventionsgruppe, n=13; Kontrollgruppe, n=12). Beide Gruppen besuchten parallel sechsmonatige Qualifizierungskurse mit 600 Unterrichtsstunden an derselben Sprachschule (DRK Akademie Schleswig-Holstein, Kiel), mit demselben Lehrbuch und demselben Curriculum. 92% der Teilnehmenden (23/25) stammten aus Tunesien; die Herkunftsländer der übrigen waren Nepal (n=1) und die Ukraine (n=1). Alle begannen das Training mit vergleichbarer Deutschkompetenz (GER-Niveau B1) und vergleichbaren demografischen Merkmalen (vgl. Tabelle 1).
Die Interventionsgruppe erhielt Zugang zu casuu (casuu GmbH, Hamburg, Deutschland, [https://casuu.com/]), einer digitalen Simulationsplattform, die interaktive, KI-gestützte Kommunikationsszenarien bereitstellt, in denen reale klinische Situationen mit virtuellen Patient*innen und Kolleg*innen simuliert werden [14]. Die Nutzung erfolgte überwiegend selbstgesteuert mit zusätzlich drei jeweils etwa zweiwöchigen Phasen, in denen casuu strukturiert in den Unterricht integriert wurde, einschließlich zugewiesener Hausaufgaben und dedizierter Debriefing-Sitzungen. Die Kontrollgruppe hatte zur selben Zeit normalen Unterricht. In den 15- bis 20-minütigen Debriefing-Sitzungen präsentierten einzelne Teilnehmende die Transkripte ihrer virtuellen Interaktionen der Klasse per Beamer. Unter Anleitung der Lehrkraft diskutierte die Klasse anschließend neu auftretende Wortfelder, Vokabellücken und die Frage, ob die verwendeten Ausdrücke bereits im Unterricht behandelt worden waren. Die Kontrollgruppe nahm an einem regulären Sprachtraining teil, das kein VI-Training enthielt, sich ansonsten jedoch hinsichtlich Curriculum und Inhalten nicht unterschied.
Parallel zu den Sprachkursen wurden die Teilnehmenden beider Gruppen auf den Stationen des Städtischen Krankenhauses Kiel (SKK) eingearbeitet. Im Verlauf des Programms verschob sich das Verhältnis von praktischer Tätigkeit zu Sprach(schul)unterricht schrittweise von etwa 1:4 Tagen pro Woche zu Beginn auf 4:1 gegen Ende. Struktur und Intensität des Sprachunterrichts sowie der praktischen klinischen Tätigkeit waren in beiden Gruppen identisch.
Primärer Endpunkt war die Durchfallquote beim ersten Versuch der berufsbezogenen Sprachprüfung B2 Pflege (telc® Deutsch B2 Pflege), die obligater Teil des Anerkennungsverfahrens in bestimmten Bundesländern ist (darunter Schleswig-Holstein; in anderen Bundesländern kann direkt eine Kenntnisprüfung abgelegt werden, die Berufssprache bleibt für deren Bestehen jedoch eine zentrale Hürde). Beide Gruppen legten diese Prüfung etwa acht Monate nach ihrer Ankunft an derselben externen Einrichtung ab (Inlingua Sprachcenter Kiel, Kiel). Das Gesamtergebnis der Prüfung wurde als sekundärer Endpunkt erfasst. Darüber hinaus wurden Trainingseffekte in verschiedenen sprachlichen Kompetenzbereichen – rezeptive Kompetenzen (Lesen und Hören), Schreiben und Sprechen – anhand der anonymisierten Ergebnisse auf Prüfungsteil-Ebene analysiert. Auch die Abbruchquoten aus dem Anerkennungsverfahren wurden erfasst. Der primäre Endpunkt wurde anschließend sowohl für alle eingeschlossenen Teilnehmenden analysiert (Intention-to-treat, ITT)

**Tabelle 1: Demografische Daten, Qualifikationsniveau und Berufserfahrung der Studiengruppen**

Für die Berufserfahrung wurden die Jahre der Tätigkeit in der Pflege im Krankenhaussetting berücksichtigt; die Daten wurden auf volle Jahre gerundet.

| | Kontrolle | Intervention | p |
|---|---|---|---|
| Alter [Jahre]: MW (Std.Abw.) | 30,2 (8,3) | 30,1 (5,1) | 0,97 |
| Geschlecht [% männlich] | 23,1 | 33,3 | 0,57 |
| BA Pflege o. gleichw. [%] | 100 | 100 | |
| Erfahrung [Jahre] | 1,23 (0,83) | 1,00 (0,00) | 0,35 |

als auch für diejenigen, die bis zur Sprachprüfung im Programm verblieben (Per-Protokoll, PP).

Um die Perspektive der klinischen Praxisanleiter*innen auf Kompetenzerwerb, Einarbeitung und Integration der Pflegefachpersonen zu untersuchen, wurden semi-strukturierte Interviews mit vier klinischen Praxisanleiter*innen am SKK durchgeführt. Jedes Interview dauerte zwischen 30 und 45 Minuten und wurde von einer einzelnen interviewenden Person durchgeführt, die die Interviews in strukturierten Zusammenfassungen dokumentierte. Vor dem Interview gaben die Teilnehmenden ihr Einverständnis zur anonymen Verarbeitung ihrer Antworten. Die thematische Struktur des Interviewleitfadens diente als deduktiver Rahmen, innerhalb dessen die Antworten der Teilnehmenden zu übergreifenden Aussagen verdichtet wurden. Dies wurde anhand der Zusammenfassung der Interviews von drei unabhängigen Leser*innen überprüft und verifiziert. Der Interviewleitfaden ist in Anhang 1 verfügbar.

## Ergebnisse

Von den Teilnehmenden, die an der Prüfung B2 Pflege wie geplant teilnahmen (PP), fielen in der Interventionsgruppe 25,0% (3 von 12) durch, verglichen mit 66,7% (6 von 9) in der Kontrollgruppe. Dies entspricht einer relativen Reduktion um 62,5% (p=0,028) in der Per-Protokoll-Analyse.

Im Verlauf des Trainings beendete eine Person aus der Interventionsgruppe ihren Vertrag mit dem SKK und brach das Anerkennungsverfahren ab. In der Kontrollgruppe war dies bei zwei Pflegefachpersonen der Fall. Zusätzlich legte in der Kontrollgruppe eine Person ihre B2-Prüfung während eines Heimaturlaubs ab und verletzte damit faktisch das Studienprotokoll. Um den Effekt der Intervention nicht zu überschätzen, wurde dieser Fall konservativ als bestanden gewertet. In der ITT-Analyse sind aus der Interventionsgruppe so 30,8% (4 von 13) als „nicht bestanden“ zu werten (Prüfung nicht bestanden oder vorzeitig ausgeschieden), verglichen mit 66,7% (8 von 12) aus der Kontrollgruppe. Dies entspricht einer relativen Reduktion der Durchfallquote um 53,8% (p=0,036).

Die Gesamtpunktzahlen in der Prüfung lagen bei 140,25±17,15 in der Interventionsgruppe und bei 125,50±14,89 in der Kontrollgruppe (p=0,030). Bei nur geringen und nicht signifikanten Unterschieden in den Kompetenzdimensionen „rezeptiv“ und „Schreiben“ war der Unterschied hauptsächlich auf „aktives Sprechen“ zurückzuführen (Arithmetisches Mittel 88,58 ± Standardabweichung 8,83 gegenüber 77,39±9,84; p=0,009).

Die Abbruchquote innerhalb der Qualifizierungsprogramme lag bei 1/13 in der VI-Gruppe gegenüber 2/12 in der Kontrollgruppe. Das Abbruchrisiko war durch das zusätzliche Training daher nicht erhöht. Die Intensität des VI-Trainings war in Phasen höher, in denen es systematisch durch strukturierte Debriefings in den Präsenzunterricht integriert wurde (62-100% wöchentlich aktive Nutzer*innen), verglichen mit anderen Wochen (0-62%).

Teilnehmende mit VI-Training wurden von ihren Praxisanleiter:innen in der klinischen Kommunikation und in zwischenmenschlichen Interaktionen als sicherer eingeschätzt. In den semi-strukturierten Interviews beschrieben alle vier Interviewten eine deutliche Verringerung der Sprachangst und eine Zunahme kommunikativer Sicherheit. Die Pflegefachpersonen wurden durchgängig als flüssiger sprechend, aktiver in Teaminteraktionen und kompetenter in patient*innennahen Situationen beschrieben. Alle Befragten stellten einen direkten Zusammenhang zwischen Sprachkompetenz und Patient*innensicherheit her. Eine höhere Sprachkompetenz in der Interventionsgruppe trug aus Sicht der klinischen Praxisanleiter*innen zudem zu einem schnelleren Einarbeitungsprozess bei.

## Schlussfolgerung und Ausblick

Das Training mit virtuellen Interaktionen (VI) kann den Erwerb berufsbezogener Sprache bei internationalem Gesundheitspersonal messbar verbessern kann. Die Integration von VI-Training erhöhte die Abbruchquoten nicht und war insbesondere im Bereich des aktiven Sprechens wirksam.

Diese Studie weist mehrere Limitationen auf: Erstens ist die Gesamtstichprobengröße begrenzt. Zweitens kann angenommen werden, dass ein größerer Effekt erzielt werden könnte, wenn das Training früher und bereits im Herkunftsland beginnt. Beides wird im Fokus einer größeren Studie stehen (V-PREP-CARE im Rahmen der Civic Innovation Plattform des Bundesministeriums für Arbeit und Soziales). Drittens war es den Teilnehmenden nicht untersagt, die Übungen gemeinsam zu bearbeiten, wodurch die Daten zur Nutzer*innenaktivität in Phasen ohne unterrichtsintegrierte Nutzung vermutlich unscharf wurden und tatsächliche Aktivität unter-, mithin den Effekt der Debriefings überschätzen könnten. Viertens kann, obwohl sie nicht aktiv informiert wurden, angenommen werden, dass die Praxisanleiter*innen wussten, welche

Teilnehmenden ein zusätzliches VI-Training absolvierten. Die Ergebnisse der semi-strukturierten Interviews können daher einem Beobachterbias unterliegen. Im Gegensatz dazu waren die Prüfenden der Prüfung B2 Pflege hinsichtlich der Gruppenzuteilung der Teilnehmenden verblindet. Fünftens spiegeln die Ergebnisse der Interviews aufgrund des zusammenfassenden Charakters der qualitativen Daten interpretierte Inhalte wider und keine wörtlich transkribierten Aussagen der Befragten. Sie sollten daher eher als explorativ denn als umfassend verstanden werden. Sechstens unterschieden die von der Prüfungsstelle berichteten anonymisierten Testergebnisse nur zwischen den Kompetenzdimensionen „rezeptiv", „Schreiben" und „aktives Sprechen", wodurch Hören und Lesen in einer einzigen Kategorie zusammengeführt wurden. Dennoch legen die Daten nahe, dass Hör- und Leseverstehen besonders relevante Zielbereiche für zukünftige Weiterentwicklungen des Trainingsprogramms sein könnten.
Insgesamt legen die Daten nahe, dass das Training mit virtuellen Interaktionen Kosten und Zeit bis zur klinischen Einsatzfähigkeit internationaler Pflegefachpersonen reduzieren kann. Als Beitrag zu einer effizienteren Arbeitsmarktintegration kann dies die Gesundheitsversorgung im Kontext von Arbeitskräftemangel und -migration unterstützen. Es sollte als potenzieller Bestandteil nachhaltiger und zukunftsfähiger Strategien zur Bewältigung des Pflegekräftemangels in Deutschland und darüber hinaus betrachtet werden.

# Anmerkungen

## Beitrag der Autor*innen

US und JMH teilen sich die Erst-Autorenschaft.
PMK und RS teilen sich die Letzt-Autorenschaft.

## ORCIDs der Autor*innen

- Jan-Marcus Haus: [0000-0003-0012-4137]
- Tom Friemel: [0009-0001-1250-9256]
- Olha Demianiuk: [0009-0000-6784-5305]
- René Werner: [0000-0002-2296-3305]
- Parisa Moll-Khosrawi:[0000-0003-2024-0020]
- Rüdiger Schmitz: [0000-0002-2414-9817]

# Danksagung

Wir danken den Pflegefachpersonen, die an dieser Studie teilgenommen haben und ihren Lehrkräften, den Mitarbeitenden der DRK Akademie SH sowie ihren Praxisanleiter*innen und den Mitarbeitenden des Städtischen Krankenhauses Kiel. Wir danken Inlingua Kiel und der telc® gGmbH für ihre Unterstützung. Wir danken Sebastian Zimmermann von der telc gGmbH für die Zusammenstellung und Anonymisierung der Prüfungsergebnisse.

# Interessenkonflikt

US, OS und SL haben Funktionen in der Folgestudie V-PREP-CARE. RW berät die casuu GmbH pro bono und ohne finanzielle Interessen. RS ist Mitgründer und Gesellschafter von casuu GmbH [https://casuu.com/]. OD ist studentische Hilfskraft im Unternehmen.
Die anderen Autor*innen erklären, dass sie keinen Interessenkonflikt im Zusammenhang mit diesem Artikel haben.

# Anhänge

Verfügbar unter https://doi.org/10.3205/zma001868

1. Anhang_1.pdf (206 KB)
   Interviewleitfaden: Praxisanleiter*innen und internationale Fachpflegekräfte

# Literatur

**Korrespondenzadresse:**

Rüdiger Schmitz
Universitätsklinikum Hamburg-Eppendorf, Institut für Angewandte Medizininformatik, Quartier am Zeughaus, Christoph-Probst-Weg 1, 20251 Hamburg, Deutschland
r.schmitz@uke.de